\documentclass[10pt]{article}
\pdfoutput =1
\usepackage{graphicx}
\usepackage{amsmath,amssymb,amsthm,latexsym}
\usepackage{amsfonts} 
\usepackage{caption}
\usepackage{subfigure}
\usepackage{slashed}
\usepackage{float} 
\DeclareGraphicsExtensions{.pdf}

\begin{document}
	\date{}
	\title{{\bf{\Large A consistent approach to the path integral formalism of quantum mechanics based on the maximum length uncertainty}}} 
	\author{{\normalsize Souvik PRAMANIK}$
		$\thanks{E-mail:  souvick.in@gmail.com}\\[0.5cm]
		{\small{\emph{Department of Mathematics, JIS University,}}}\\
		{\small{\emph{81, Nilgunj Rd, Jagarata Pally, Deshpriya Nagar, Agarpara, Kolkata 700109, West Bengal, India }}}}
	
	
	\maketitle
	\begin{abstract}
		We have developed a proper path integral formalism consistent with the deformed version of the quantum mechanics which contains a maximum observable length scale at the order of the Cosmological particle horizon, existing in cosmology. First, we have presented the modifications to the classical mechanics which shows  non-minimal effects on the equation of motion of a particle. Next, we have provided representation of the deformed quantum mechanical algebra. With this algebra in hand, we have calculated the general form of the path integral propagator in this deformed background. Finally, as a most simple case, we have built up the explicit form of the free particle propagator. The modifications to the free particle propagator shows some non-trivial effects in this case, which can have some important significance.\\
		

\end{abstract}

\section{Introduction}

	It is known that Quantum Mechanics is a well establish theory to explain all microphysical systems and those are proven to be true through a lots of experiments. In this sense, it is a self-consistent theory. However, the quantum mechanics is not compatible with the general theory of relativity due to the nonrenormalizable divergence appears whenever one tries to quantize gravity at quantum level. This is one of the main problem to build up a theory of quantum gravity that has been going on for a long time. Although, the String theory is capable to combine these two, but it has not yet been possible to verify the quantum gravity scenarios, experimentally.  So, one should think of some alternative theories. In fact, such an alternative theory has been proposed which is basically some modifications of the quantum mechanics \cite{PRD.95.103523.2017}. This theory is motivated to explain the quantum gravity phenomena within our observable universe. In fact, the length of the observable universe has been considered to be the maximum position uncertainty  one can obtained through an experiment. Such a generalization of the quantum mechanics is viable and can have interesting prospects \cite{PRD.100.12.123527.2019, EPL.135.59001.2021}. We already knows a quantum theory that has a minimum length scale as its lower bound, beyond which one can not probe \cite{PLB.304.65.1993}. For that, the quantum mechanics has been modified according to the Generalized Uncertainty Principle (GUP) \cite{PRD.52.1108.1995,JPA.30.2093.1997,PRL.101.221301.2008,IJGMMPD.23.1430025.2014}, which has been proposed from many approaches to the quantum gravity, namely, the String theory, the Double special relativity theory and the Black hole physics. Those are basically quantum gravity phenomenology based on some thought experiments, which shows that, at a very high energy regime, a theoretical cutoff scale of energy, or equivalently, at a very small scale, a cutoff scale of length mast be included into the theory \cite{IJGMMPD.23.1430025.2014}. And so, the Heisenberg uncertainty principle is replaced by the GUP. The existence of a minimum measurable length scale or fundamental ultraviolet cutoff scale at the order the Planck length is cordially associated with the GUP. Later on, a most general form of GUP has been proposed which incorporate both the minimum measurable length and maximum measurable momentum \cite{PLB.678.497.2009}. However, the phenomenology here depends on the maximum observable length scale which is our point of interest \cite{PRD.95.103523.2017}. We known that the quantum mechanics is a non-local theory and the cosmological particle horizon is the maximum measurable length of our Universe. So, in order to study the global properties of the Universe by considering quantum mechanics as a global theory, the cosmological particle horizon has been incorporated in to the quantum mechanical uncertainty relation as a maximum length uncertainty. It is assumed that the global properties of the Universe could produce a signature on the behavior of local quantum systems. Inspired by this agenda, we want to study the properties of the modified quantum mechanics according to this model. First, we want to build up the representation of the algebra and study the symmetries. Afterward, we will develop a general path integral formalism with the modified algebra in hand. We will explicitly show the modification to the free particle propagator and study its consistency. Since, we have some deformed uncertainty relation which interns affects the quantum commutators, we will derive everything in the deformed representation. In the next section, we will show how to jump to deformed variables (non-canonical) from conventional one. Thereafter, we will study the modifications to the Classical Mechanics corresponding to the deformed algebra followed by representation of the quantum mechanical algebra. In the forth section, we will develop the modified path integral formalism in the deformed background. There we will also calculate the free particle propagator explicitly and show its consistency. We will conclude our findings in the conclusions and future prospects  section.

	\section{Noncanonical Hamiltonian dynamics}
	
	Suppose we have a system of ordinary differential equations of the form
	\begin{equation}
	\frac{d z_i}{dt} = P_i(z), \ \ \ i=1,2,...,N,
	\end{equation}
	where, the $z_i$'s are the phase space variables. First, how would we know if the above system is a Hamiltonian system or not? One way to check this is to consider Liouville's theorem, which tells that the Hamilton's equations have the property $\partial \dot x_i / \partial x_i
	+ \partial \dot p_i / \partial p_i = \partial^2 H / \partial x_i \partial p_i - \partial^2 H / \partial p_i \partial x_i = 0$. However, for the system $\dot z_1 = - z^2_2/3 -z_1, \dot{z}_2 = z_1/ (z_2^2 + 1)$, we have $\partial \dot x_i / \partial x_i
	+ \partial \dot p_i / \partial p_i \neq 0 $. But, the above system is a disguised simple harmonic oscillator. It has been disguised by making a non-canonical coordinate change \cite{RMP.70.2.1998}. So, whenever the non-canonical representations are concerned, Liouville's theorem is unable to determine whether a system is a Hamiltonian system or not. Indeed, there does not exist any specific  way to check whether a system is Hamiltonian or not in the non-canonical representation. However, here we can specify a technique to jump to non-canonical variables as follows - first consider a Hamiltonian system which has a finite number of degrees of freedom in the phase space in canonical representation 
	\begin{equation} 
	z^i_c= {\Huge\{ } {\begin{array}{c}
		x^i \ \ for \ \  i=1,2,...,N, \\
		p^{i-N} \ \ for \ \ i=N+1,...,2N. \\
		\end{array} } 
	\end{equation}
      In terms of them, the Hamilton's equations obtain the covariant form as, $\dot z^i_c = J^{ij}_c \frac{\partial H_c}{\partial z^j_c} = [z^i_c,H_c]$, where $J^{ij}_c$ is the squire matrix of $N\times N$ blocks 
	\begin{equation} J^{ij}_c = 
	\left[ {\begin{array}{cc}
		0_N & I_N \\
		I_N & 0_N \\
		\end{array} } \right]
	\end{equation}
	which is associated with the Poisson brackets in the manner $[A,B]_c = \frac{\partial A}{\partial z^i}J^{ij}_c\frac{\partial B}{\partial z^j}$ for $i,j$ run over $1,2, . . . ,2N$. Now, if we can consider a general non-canonical change of variables $z^i_{nc} \equiv z^i_{nc}(z_c)$, then the Hamilton's equations of motion undergo the changes as
	$\dot z^i_{nc} = (\frac{\partial z^i_{nc}}{\partial z^j_c} J^{jk}_c \frac{\partial z^l_{nc}}{\partial z^k_c}) \frac{\partial H_{nc}}{\partial z^l_{nc}}$. If we define $J^{il}_{nc} = \frac{\partial z^i_{nc}}{\partial z^j} J^{jk}_c \frac{\partial z^l_{nc}}{\partial z^k}$, then the  Hamilton's equations of motion remain form independent in the non-canonical representation
	\begin{equation}
	\dot z^i_{nc} = J^{il}_{nc} \frac{\partial H_{nc}}{\partial z^l_{nc}} = [z^i_{nc},H_{nc}]
	\end{equation}
	where, the new Poisson bracket is defined as $[A,B]_{nc} = \frac{A}{z^i_{nc}}J^{il}_{nc}\frac{B}{\partial z^l}$.
	The above Hamilton's equations can be inverted to 
	$(J^{il}_{nc})^{-1} \dot z_{nc}^i = \frac{\partial H_{nc}}{\partial z^l_{nc}}$.
	We want to achieve this equation from the variation of the action 
	\begin{equation}
	I=\int [a^i(z_{nc}). \dot{(z_{nc})_i} - H(z_{nc}) ] dt \label{action 1}
	\end{equation}
	with respect to the variable $z_{nc}$ \cite{PLA.380.20.1714.2016}. This can be done only if the following equation is satisfied
	\begin{equation}
	\frac{\partial a^j}{\partial (z_{nc})_i} - \frac{\partial a^i}{\partial (z_{nc})_j} = (J^{ij}_{nc})^{-1}. \label{ai-j relation}
	\end{equation}
	In the next section, we will explicitly calculate the form of the matrix $J^{ij}_{nc}$ and the action (\ref{action 1}) for our model of interest.
	
	\section{Maximal length Uncertainty relation: deformed algebra and representations}
	
	The one dimensional maximal length uncertainty principle has been proposed to be \cite{PRD.95.103523.2017}
	\begin{equation}
	\Delta x.\Delta p \geq \frac{\hbar}{2}\frac{1}{1- \beta \Delta x^2}, \label{max-gup}
	\end{equation}
	where, $\beta = l_{max}^{-2} \simeq (H_0/c)^2$ ($H_0$ is the Hubble parameter and c is the speed of light). Beside the consideration of a maximum measurable length scale $l_{max}=1/  \sqrt{\beta}$, this form of uncertainty relation also implies a minimum measurable momentum, such as $p_{min}= 3 \sqrt{3} /4 \times \hbar \sqrt{\beta}$. The quantum mechanical commutation relations consistent with the above form of uncertainty relation are
	\begin{equation}
	[\hat x, \hat p] = i \hbar \frac{1}{1 - \beta x^2}, \ [\hat x, \hat x] = 0. \ [\hat p,\hat p] =0. 
	\end{equation}

\subsection{Classical counterpart}

The above commutation relations forces the Poisson bracket structures in classical mechanics to be a non-canonical one like 
	\begin{equation}
	\{ x, p \} = \frac{1}{1 - \beta x^2}, \ \{ x,  x\} = 0, \ \{ p, p \} =0. \label{poisson}
	\end{equation}
	Though the above structure is completely non-canonical, however a route to the canonical variables $(X,P)$ can be made as
	\begin{equation}
	x =  X, \ \ \ p = \frac{P}{1 - \beta X^2}. \label{ca-non-ca}
	\end{equation}
    Looking at (\ref{poisson}), we can write the matrix $J_{ij}$ (since we are in non-canonical representation, we will just write $J_{ij}$ instead of $(J_{nc})_{ij}$) in this case as
     \begin{equation}
     J_{ij} = \left[{\begin{array}{cc}
     	\{ x,  x\} & \{ x, p \}\\
     	\{ p, x \} & \{ p, p \}\\
     	\end{array}}\right] = \left[{\begin{array}{cc}
     	0 & \frac{1}{1 -  \beta x^2}\\
     	-\frac{1}{1 -  \beta x^2} & 0\\
     	\end{array}}\right].
     \end{equation} 
     The above block matrix can be inverted to get
     \begin{equation}
     (J^{ij})^{-1} = \left[{\begin{array}{cc}
     	0 & -1 + \beta x^2\\
     	1 - \beta x^2 & 0\\
     	\end{array}}\right]. \label{j-inv}
     \end{equation}
     The above matrix in comparison with (\ref{ai-j relation}) provides two sets of structures for the functions $a_x$ and $a_p$, as  
     \begin{equation}
     a_x = 0, \ a_p = -x + \frac{\beta x^3}{3}, \ \ (Or)\ \ a_x =  p (1 - \beta x^2), \ a_p = 0.
     \end{equation}
     So, the deformed structures of the action consistent with the above model (\ref{poisson}) turns out to be
     \begin{equation}
     I = \int \left[- \left(x - \frac{\beta x^3}{3}\right)\dot{p} - H(x,p)\right] dt = \int \left[p (1 - \beta x^2) \dot{x} - H(x,p)\right] dt. \label{action 2}
     \end{equation}
     The two forms of the above action are equivalent in a sense that one can be converted to the other just by performing integration by parts. In this sense, whichever of them is considered, they will end up with the same dynamics. If we consider the Hamiltonian $H(x,p) = p^2/2m +V(x)$, then the Hamilton's equation of motion 
     $$ \dot x = [x,H] , \ \ \ \dot p = [p,H]$$
     provides the equation of motion of the particle by using the above Poisson brackets (\ref{poisson}) to be
     \begin{equation}
     m((1 - \beta x^2)^2 \ddot x - 2 \beta x (1 - \beta x^2) \dot x^2)+ V_x(x)= 0. \label{eom}
     \end{equation}
     In order to maintain the consistency, we want to obtain the above equation of motion from a consistent Lagrangian system. Looking at the above equation one may guess that the corresponding Lagrangian must contain a non-minimal coupling of $x$ and $\dot x$. In fact, if we assume the form of the Lagrangian as $L = 1/2m \ \dot x^2 \times f(x) - V(x)$, then it is straightforward to obtain the above equation, if $f(x) = (1 - \beta x^2)^2$. This evaluates the form of the Lagrangian compatible with (\ref{eom}), is
     \begin{equation}
     L(x, \dot x, t) = \frac{1}{2m} \dot x^2 (1 - \beta x^2)^2 - V(x). \label{lag}
     \end{equation} 
     The above action (\ref{action 2}) can be obtained from the above Lagrangian (\ref{lag}), or, in other words consistent with (\ref{lag}), if we define the non-canonical momentum to be $p = \frac{\partial L}{\partial [\dot x (1 - \beta x^2)]} = m \dot x (1 - \beta x^2)$. It should be pointed out here that this non-canonical momentum cannot be obtained from the equation (\ref{ca-non-ca}) above. So, if someone tries to calculate the momentum from equation (\ref{ca-non-ca}), it will be inconsistent. With (\ref{eom}) in hand, if we solve the equation for a free particle ($V(x) = 0$), we will obtain $x(t) = t + \beta t^3/3$ to the first order of $\mathcal{O}(\beta)$. So, the trajectory no longer will be a straight line.  
         	
	\subsection{Representation of algebra}
	
	In the coordinate representation, we define the condition for completeness of the eigenvectors $| x \rangle$ as
	\begin{equation}
	\mathbb{I} = \int_{-\infty}^{\infty} |x\rangle \langle x| (1- \beta x^2) dx, \label{identity-co} 
	\end{equation}  
	and the inner product between two basis vector as $\langle x | x'\rangle = \frac{\delta(x - x')}{(1 - \beta x^2)}$. So, the inner product between any two vectors can be calculated to be
	\begin{equation}
	\langle \psi | \psi'\rangle = \langle \psi |\mathbb{I} | \psi'\rangle = \int_{-\infty}^{\infty} \langle\psi  |x \rangle \langle x| \psi' \rangle (1 - \beta x^2) \ dx = \int_{-\infty}^{\infty} \psi^*(x) \ \psi'(x) \  (1 - \beta x^2) \ dx \label{inner-pro-co}
	\end{equation}
	Though the co-ordinate representation has been modified, but we will remain the momentum representation unaltered so that the following relations hold
	\begin{eqnarray}
	&&\mathbb{I} = \int_{-\infty}^{\infty} |p\rangle \langle p| dp , \ \ with \ \  \langle p | p'\rangle = \delta(p - p') \label{identity-mo} \nonumber \\ 
    && \langle \psi | \psi'\rangle = \langle \psi |\mathbb{I} | \psi'\rangle = \int_{-\infty}^{\infty} \langle\psi  |p \rangle \langle p| \psi' \rangle  \ dp = \int_{-\infty}^{\infty} \psi^*(p) \ \psi'(p) \   \ dp. \label{inner-pro-mo}
	\end{eqnarray}
	As usual if we define the web function as $\psi(p) = \langle p | \psi \rangle$ and $\psi(x) = \langle x | \psi \rangle$, the set of position and the momentum operators acting on web function in coordinate and momentum representation are
	\begin{eqnarray}
	&& \hat x \psi(x) = x \psi(x), \ \hat p \psi(x) = - \mathrm{i} \hbar \frac{1}{1 - \beta x^2} \frac{\partial \psi (x)}{\partial x}\nonumber\\
	&& \hat x \ \psi(p) = \frac{1}{\sqrt{\beta}} \tanh^{-1} \left( \mathrm{i} \hbar \sqrt{\beta} \frac{\partial}{\partial p} \right) \ \psi (p), \ \ \hat p \psi(p) = p \psi(p) \label{po-mo}
	\end{eqnarray}
	Both the above representations shows the inner product between the eigenstates $|x\rangle$ and $| p \rangle$ can be calculated to be \begin{equation}
	\langle x | p \rangle \propto  e^{\frac{i p}{\hbar} \left(x - \beta \frac{x^3}{3} \right)} \label{in-co-mo}
	\end{equation} 
up tp the first order of $\mathcal{O}(\beta)$. This inner product is consistent with the classical action (\ref{action 2}) and therefore will be used to construct the path integral propagator. The above position and momentum operators (\ref{po-mo}) obey the symmetries $ (\langle \psi | \hat x)| \psi' \rangle = \langle \psi |( \hat x| \psi' \rangle) , \ (\langle \psi | \hat p)| \psi' \rangle = \langle \psi |( \hat p| \psi' \rangle)$ in both the co-ordinate and momentum representations. This which can be shown as follows: 
	In the coordinate representation 
	\begin{eqnarray}
	(\langle \psi |\hat x) | \psi' \rangle &=& \int_{-\infty}^{\infty} \int_{-\infty}^{\infty}  ( \langle \psi | x_1 \rangle \langle x_1|  \hat x ) | x_2 \rangle \langle x_2 | \psi' \rangle (1 - \beta x_1^2) (1 - \beta x_2^2) dx_1 \ dx_2 \nonumber\\
	&=& \int_{-\infty}^{\infty}  x_1 \psi^*(x_1) \psi'(x_1) (1 - \beta x_1^2) dx_1 =  \langle \psi |( \hat x| \psi' \rangle). \nonumber \\
	(\langle \psi |\hat p) | \psi' \rangle &=& \int_{-\infty}^{\infty} \int_{-\infty}^{\infty}  ( \langle \psi | x_1 \rangle \langle x_1|  \hat p ) | x_2 \rangle \langle x_2 | \psi' \rangle (1 - \beta x_1^2) (1 - \beta x_2^2) dx_1 \ dx_2 \nonumber\\
	&=& \int_{-\infty}^{\infty} \int_{-\infty}^{\infty}  \left[ - \i\hbar \frac{\partial \psi^*(x_1)}{\partial x_1} \psi'(x_2) \frac{(1- \beta x_2^2)}{(1 - \beta x_1^2)}\delta(x_1-x_2)dx_1 dx_2\right]\nonumber\\ 
	&=& \int_{-\infty}^{\infty} - \i\hbar \frac{\partial \psi^*(x_1)}{\partial x_1} \psi'(x_1) dx_1 = \int_{-\infty}^{\infty} \psi^*(x_1) i\hbar \frac{\partial \psi'(x_1)}{\partial x_1} dx_1 = \langle \psi |( \hat x| \psi' \rangle).
	\end{eqnarray}
	In momentum representation, the momentum operator  $\hat p$ as usual satisfies the symmetry condition. For the position operator the symmetry can be shown as
	
	\begin{eqnarray}
	(\langle \psi |\hat x) | \psi' \rangle &=& \int_{-\infty}^{\infty} \int_{-\infty}^{\infty}  ( \langle \psi | p_1 \rangle \langle p_1|  \hat x ) | p_2 \rangle \langle p_2 | \psi' \rangle dp_1 dp_2 \nonumber\\
	&=& \int_{-\infty}^{\infty} \int_{-\infty}^{\infty}   \frac{1}{\sqrt{\beta}} \left[ \tanh^{-1} \left(i\hbar \frac{\partial}{\partial p_1}\right)\right]\psi^*(p_1)\delta(p_1-p_2) \psi'(p_2) dp_1 dp_2 \nonumber\\ 
	&=& \int_{-\infty}^{\infty} \int_{-\infty}^{\infty}   \frac{1}{\sqrt{\beta}} \sum_{j=0}^\infty \frac{1}{(2j+1)}\left(i\hbar \frac{\partial}{\partial p_{1}}\right)^{2j+1} \psi^*(p_1)\delta(p_1-p_2) \psi'(p_2) dp_1 dp_2 \nonumber\\
	&=& \int_{-\infty}^{\infty} \int_{-\infty}^{\infty}   -\psi^*(p_1) \frac{1}{\sqrt{\beta}} \sum_{j=0}^\infty \frac{1}{(2j+1)}\left(i\hbar \frac{\partial}{\partial p_{1}}\right)^{2j+1} \psi'(p_1) dp_1 \nonumber\\
	&=&  \langle \psi |( \hat x| \psi' \rangle).
	\end{eqnarray} 
	This is the representation of the algebra consistent with the deformed uncertainty principle (\ref{max-gup}).  
	
	\section{Path integral in one dimension}
	
	In this section, first we should point out that there are two ways to achieve any effects in the non-canonical representation other than the usual canonical representation. Either one has to consider the usual Hamiltonian with the deformed algebras or one can consider a deformed Hamiltonian induced by deformed variables with conventional algebras. Else, everything will be just a variable transformation from the canonical to the non-canonical one through equation (\ref{ca-non-ca}). Here, we will consider the earlier, which is indeed consistent with the action (\ref{action 2}). So, the form of the Hamiltonian, we chose is
	\begin{eqnarray}
	H = \frac{\hat p^2 }{2m} + V(\hat x). \label{ham} 
	\end{eqnarray}
	The Schrodinger's equation in the momentum and coordinate representations can be achieved using (\ref{po-mo}), as
	\begin{equation}
	\left[ \frac{p^2}{2m} + V\left(\frac{1}{\sqrt{\beta}} \tanh^{-1}\left(\mathrm{i} \hbar \sqrt{\beta} \frac{\partial}{\partial p}\right)\right)\right]  \psi (p,t) = i \hbar \frac{\partial \psi (p,t)}{\partial t}, \label{sch-mo}
	\end{equation}
	\begin{equation}
	\left[ - \frac{\hbar^2}{2m}\left( \frac{1}{(1 - \beta x^2)^2}\partial_x^2 + \frac{2 \beta x}{(1 - \beta x^2)^3}\partial_x \right) + V\left(x\right)\right] \psi (x,t) = i \hbar \frac{\partial \psi (x,t)}{\partial t}. \label{sch-co}
	\end{equation}
If a quantum mechanical system undergo a displacement from the quantum state $x'$ at time $t'$ to any later state $x''$ at time $t''$, then the one dimensional quantum mechanical path integral propagator in momentum representation is $K(x'',t'',x',t') = \langle x'', t'' | x', t' \rangle = \langle x''| e^{- \frac{i}{\hbar } H (t'' - t')}| x' \rangle$. We divide the time interval $t'' - t'$ into $N$ subintervals of equal length $\epsilon$, so that $t' =t_0$, $t_i = t_0 + i \epsilon$, $t_N = t''$. If $x_i = x(t_i)$'s are the position at the time $t_i$, then applying the completeness relation (\ref{identity-co}) for each intermediate time $t_i$, we have the propagator
	\begin{equation}
	K(x'',t'',x',t') = \int \langle x''| e^{- \frac{i}{\hbar } H \epsilon}| x_{N-1} \rangle \langle x_{N-1}| e^{- \frac{i}{\hbar } H \epsilon}| x_{N-2} \rangle...\langle x_1| e^{- \frac{i}{\hbar } H \epsilon}| x' \rangle \prod_{i=1}^{N-1}(1 - \beta x_i^2) dx_i.\label{prop - full}
	\end{equation}
In order to calculate the propagator $\langle x_{i}| e^{- \frac{i}{\hbar } H \epsilon}| x_{i-1} \rangle$ for the short time interval $(t_i - t_{i-1})$, it is reasonable to use the Weyl ordered form of the exponential function $(e^{-i\epsilon /\hbar H })_{WO}$ in the momentum representation. The benefit to use a Weyl ordered form is that, any product of the operators $\hat x^n$ and $\hat p^m$, for any $n, \ m \in \mathbb{Z}$, can be written in a symmetric form. The exponential function can be expanded to give 
	$$(e^{- \frac{i}{\hbar } H \epsilon})_{WO} = \sum_{N=0}^{\infty} \frac{1}{N !} \times \left(-\frac{i \epsilon}{\hbar}\right)^N \times \left( \frac{\hat p^2}{2 m} + V(\hat x)\right)_{WO}^N.$$
The last term of the above expansion can be further expanded in terms of Weyl ordering\footnote{($(\hat x^n \hat p^m)_{WO} =  \frac{1}{2^m}\sum_{l=0}^{m} \frac{m!}{l! \ (m-l)!} \hat p^l \hat x^n \hat p^{m-l}$)}  as
	
	\begin{eqnarray}
	\left( \frac{\hat p^2}{2 m} + V(\hat x)\right)_{WO}^N
	&=& \sum_{i=0}^{N} \frac{N!}{ i! \ (N-i)!} \times \frac{1}{ \ 2^{N-i} }\sum_{l=0}^{N-i} \frac{(N-i) !}{ l! \ (N-i - l)!}  \ V(\hat x)^{N-i-l} \left(\frac{\hat p^2}{2m}\right)^i \ V(\hat x)^{l}.   
	\end{eqnarray}
Substituting all, the propagator $\langle x_{i}| e^{- \frac{i}{\hbar } H \epsilon}| x_{i-1} \rangle$ for a short time interval $(t_i - t_{i-1})$ can be written as
	\begin{eqnarray}
	&& \langle x_{i}| (e^{- \frac{i}{\hbar } H \epsilon})_{WO}| x_{i-1} \rangle \nonumber\\
	&& = \int \sum_{N=0}^{\infty} \frac{1}{N !} \left(-\frac{i \epsilon}{\hbar}\right)^N \sum_{i=0}^{N} \frac{N!}{ i! \ (N-i)!} \times \frac{1}{ \ 2^{N-i} }\sum_{l=0}^{N-i} \frac{(N-i) !}{ l! \ (N-i - l)!}  \ \langle x_i | V(\hat x)^{N-i-l}  | p \rangle \langle p | \left(\frac{\hat p^2}{2m}\right)^i  V(\hat x)^{l} | x_{i-1} \rangle \frac{dp_i}{2 \pi \hbar} \nonumber\\
	&& = \int \sum_{N=0}^{\infty} \frac{1}{N !} \left(-\frac{i \epsilon}{\hbar}\right)^N \sum_{i=0}^{N} \frac{N!}{ i! \ (N-i)!} \frac{1}{ 2^{N-i} } \nonumber\\
	&& \ \ \ \ \sum_{l=0}^{N-i} \frac{(N-i) !}{ l! \ (N - i - l)!}  \ V \left( x_i \right)^{N-i -l}  \left(\frac{p^2}{2m}\right)^i V \left( x_{i-1} \right)^{l} e^{\frac{i p}{\hbar}\left[ x_i - \beta \frac{x_i^3}{3} - x_{i-1} - \beta \frac{x_{i-1}^3}{3} \right]} \frac{dp_i}{2 \pi \hbar} \nonumber\\
	&& = \int e^{\frac{i p}{\hbar}\left[ x_i - \beta \frac{x_i^3}{3} - x_{i-1} - \beta \frac{x_{i-1}^3}{3} \right]} e^{-\frac{i \epsilon}{\hbar} \left[\frac{p^2}{2m} + V\left(\frac{x_i + x_{i-1}}{2} \right) \right]} \frac{dp_i}{2 \pi \hbar}.  \label{prop-short} 
	\end{eqnarray}
In the above, we have used the inner product	rule (\ref{in-co-mo}). 	Substituting (\ref{prop-short}) into (\ref{prop - full}) for each subinterval $(t_i -t_{i-1})$, $i = 1,2,..., N$, we obtain the propagator for any finite time interval $t'' - t'$, as
	\begin{equation}
	K(x'',t'',x',t') = \int \int  e^{ \sum_{i=1}^{N} \left\{ \frac{i p_i}{\hbar}\left[ x_i - \beta \frac{x_i^3}{3} - x_{i-1} - \beta \frac{x_{i-1}^3}{3} \right] -\frac{i \epsilon}{\hbar} \left[\frac{p_i^2}{2m} + V\left(\frac{x_i + x_{i-1}}{2} \right) \right] \right\}} \prod_{i=1}^{N-1} (1 - \beta x_i^2)  dx_i \  \prod_{i=1}^{N} \frac{dp_i}{2 \pi \hbar}  \label{pro}
	\end{equation}
If we consider the limits $N \rightarrow \infty$ with $\epsilon \rightarrow 0$, we have the approximation $\left[ x_i - \beta \frac{x_i^3}{3} - x_{i-1} - \beta \frac{x_{i-1}^3}{3} \right] \sim \epsilon \times \dot x_i (1- \beta x_i^2)$. With this, we obtain the general form of the propagator in this deformed representation
\begin{equation}
	K(x'',t'',x',t') = \int \int e^{ \sum_{i=1}^{N} \int_{t_{i-1}}^{t_i}\frac{i}{\hbar}\left\{  (p_i . \dot x_i)(1 - \beta x_i^2) -\left[\frac{p_i^2}{2m} + V\left(x_i \right) \right] \right\}dt} \prod_{i=1}^{N-1} (1 - \beta x_i^2)dx_i \prod_{i=1}^{N} \frac{dp_i}{2\pi \hbar}. \label{pro-full}
\end{equation}

\subsection{Free particle solution and the quantum mechanical propagator}

The Schrodinger equation for the free particle can be achieved by substituting $V(x) = 0 $ in (\ref{sch-co}), which leads to
\begin{equation}
- \frac{\hbar^2}{2m}\left[ \frac{1}{(1 - \beta x^2)^2}\partial_x^2 + \frac{2 \beta x}{(1 - \beta x^2)^3}\partial_x\right]  \psi (x,t) = i \hbar \frac{\partial \psi (x,t)}{\partial t} = E \ \ (Say). \label{sch-free}
\end{equation}
A separation of variable $\psi(x,t) = \psi(x) \phi(t)$ provides the special part of the above, is
\begin{equation}
\psi'' (x)+ \frac{2 \beta x}{(1 - \beta x^2)} \psi'(x) + \frac{2 m E}{\hbar^2}(1 - \beta x^2)^2 \psi (x)=0 \label{sch-free-1}
\end{equation} 
This equation has the solution 
$$ \psi(x) = A e^{ i a (x - \frac{\beta x^3}{3})} + B e^{ - i a (x -  \frac{\beta x^3}{3})}, \ \ \ a= \sqrt{2mE}/\hbar,$$ 
which leads to the free particle web function in this scenario, as
\begin{equation}
\psi(x,t) = A e^{ i a (x - \frac{\beta x^3}{3}) - \frac{i E t}{\hbar}} + B e^{ - i a (x - \frac{\beta x^3}{3}) - \frac{i E t}{\hbar}}. \label{sch-sol}
\end{equation}
It should be noted that the above solution tends to the usual propagator as $\beta \rightarrow 0$. We now concentrate to calculate the free particle propagator.   

It can be realized form  (\ref{pro}), it is quite difficult to calculate the propagator conventionally due to the presence of the factor  $(1-\beta x^2/3)$ multiplied with $(p.x)$. However, this difficulty can be cured by means of the variable transformation from  $(x,p)$ to $(a,b)$, following 
	\begin{equation}
	a = (x - \beta x^3/3), \ b = p.
	\end{equation}
In terms of $(a,b)$, the above propagator for a free particle ($V(x) = 0$) becomes
	\begin{eqnarray}
	K(a'',t'',a',t') &=& \int \int  e^{ \sum_{i=1}^{N} \left\{ \frac{i  b_i(a_i - a_{i-1})}{\hbar} -\frac{i \epsilon}{\hbar} \left[\frac{b_i^2}{2m} \right] \right\}} \prod_{i=1}^{N-1}  da_i \  \prod_{i=1}^{N} \frac{db_i}{2 \pi \hbar}.
	\end{eqnarray}
This is same as the conventional one, and therefore can easily be solved to obtain $K(a'',t'',a',t')= \sqrt{\frac{m}{2 \pi i \hbar (t'' - t')}} e^{\frac{i m (a'' - a')^2}{2 \hbar (t'' - t')}}$.	Returning back to the original variables, we have the form of the propagator to the first order of $\mathcal{O}(\beta)$, as
	\begin{eqnarray}
	K(x'',t'',x',t') &=&  \sqrt{\frac{m}{2 \pi i \hbar (t'' - t')}} e^{\frac{i }{2 \hbar (t'' - t')}\left(x'' - \beta \frac{x''^3}{3} - x' + \beta \frac{x'^3}{3} \right)^2}\nonumber\\
	&=& \sqrt{\frac{m}{2 \pi i \hbar (t'' - t')}}e^{\frac{i (x''-x')^2 }{2 \hbar (t'' - t')}\left(1- \frac{2\beta}{3}(x''^2+x''x'+x'^2) \right)} .\label{pro-final}
	\end{eqnarray}
In the structure of the free particle propagator above, it can be noticed that a non-minimal combination of initial and final conditions has emerged.

\subsection{Consistency of the propagator}
	
If $\psi(x',t') = \langle x',t'| \psi \rangle$ is the webfunction of a system at the initial state $| x',t' \rangle$, then the propagator transforms the webfunction from an initial state $| x',t' \rangle$ to a final state $| x'',t'' \rangle$ by means of the relation
	\begin{eqnarray}
	\psi(x'',t'') = \langle x'',t''| \psi \rangle &=& \int \langle x'',t''| x',t' \rangle \langle x',t' | \psi \rangle (1 - \beta x'^2) dx' \nonumber\\
	&=& \int K(x'',t'';x',t') \ \psi(x',t') \ (1- \beta x'^2) \ dx'. \label{pro-criteria}
	\end{eqnarray} 
Substituting the relations (\ref{sch-sol}) and (\ref{pro-final})  into the right hand side of (\ref{pro-criteria}), and then integrating out the initial state one can obtain the same form of the  webfunction at the final state. This reveals the consistency of the propagator (\ref{pro-final}). Looking at (\ref{pro-criteria}), one may guess that the normalization criteria of the propagator $K(x'',t'';x',t')$ will also be modified. This criteria can be shown by considering the invariance of total probability amplitude of the web function at any arbitrary time $t$. Total probability amplitudes of the web function $\psi$ at the time $t'$ and $t''$ are 
	\begin{eqnarray}
	\langle \psi| \psi \rangle_{(x',t')} &=& \int \langle \psi |x',t' \rangle \langle x',t' | \psi \rangle  (1 - \beta x'^2) dx' = \int \psi^*(x',t') \psi(x',t') (1 - \beta x'^2) dx' \label{norm - web - p't'}\\
	\langle \psi| \psi \rangle_{(x'',t'')} &=& \int \langle \psi |x'',t'' \rangle \langle x'',t'' | \psi \rangle  (1 - \beta x''^2) dx' = \int \psi^*(x'',t'') \psi(x'',t'') (1 - \beta x''^2) dx''  \label{norm - web - p''t''} 
	\end{eqnarray}
So, we have  $\langle \psi| \psi \rangle_{(x',t')} = \langle \psi| \psi \rangle_{(x'',t'')}$. Substituting the propagation criteria (\ref{pro-criteria}) into the r.h.s. of (\ref{norm - web - p''t''}) and then compare with (\ref{norm - web - p't'}), we have the normalization condition for the propagator
\begin{equation}
	\int \int K^*(x''t'';x',t') K(x'',t'',x_1,t_1) (1 - \beta x_1^2) (1 - \beta x''^2) \psi(x_1,t_1) dx'' dx_1 = \psi(x',t'). \label{norm - criteria}
\end{equation}  
Again, substituting the relations (\ref{sch-sol}) and (\ref{pro-final})  into the left hand side of the above expression one can obtain the same webfunction  (\ref{sch-sol}) at the state $(x',t')$. This shown the consistency of the above propagator. 

\section{Conclusion and future prospects } 

We have developed a proper non-canonical formulation to study the modified dynamics consistent with that deformed algebra has came from the proposed version of the quantum mechanics, in which the cosmological particle horizon existing in cosmology is the maximum measurable length uncertainty. First, we have developed the classical counterpart consistent with the deformed algebra and found the equation of motion of a particle influenced by any arbitrary potential. Indeed, we have shown both the phase space and coordinate space formalism. Looking at the Lagrangian (\ref{lag}) one can observe that the kinetic term enfluences non-minimal couplings of position coordinate which in turn affects the equation of motion (\ref{eom}). That's why, in the case of just a free particle, the trajectory will not be just a straight line, rather a curved line. Next, we have provided representation of the quantum mechanical algebra. In particular, we have built up the modified position and momentum operators and shown that they are also hermitian in this case. Thereafter, we proceeded to find the path integral propagator. In this context, we have adopted the Hamiltonian path integral formalism to find the propagator and also taken care of Weyl ordering rules. In (\ref{pro-full}), one can see the non-trivial changes occur in the quantum mechanical path integral propagator. As a most simple case, we have calculated the explicit form of the free particle propagator in this non-canonical scenario. The extra terms that have emerged in the exponential of equation (\ref{pro-final}), comparing them with the conventional case, suggest the same thing that the trajectory will no longer be a straight line in the case of just a free particle. We have further shown the consistency of the free particle propagator. With the above propagator (\ref{pro-full}) in hand and following out technique, one can observe the changes by studying other quantum mechanical systems as well \cite{PRL.101.221301.2008}.

\vskip .5cm
	
{\small {\bf{Acknowledgements:}} We would like to thank the Indian Statistical Institute and the JIS University for the support.}

\end{document}